# Strain rate dependency of dislocation plasticity

Haidong Fan[ab*], Jaafar A. El-Awady[c], Qingyuan Wang[a*], Dierk Raabe[b], Michael Zaiser[d]

[a] Department of Mechanics, Sichuan University, Chengdu 610065, China

[b] Department Microstructure Physics and Alloy Design, Max-Planck-Institut für Eisenforschung GmbH, Düsseldorf 40237, Germany

[c] Department of Mechanical Engineering, Whiting School of Engineering, The Johns Hopkins University, Baltimore, MD 21218, USA

[d] WW8-Materials Simulation, Department of Materials Science, FAU Universität Erlangen-Nürnberg, Fürth 90762, Germany

*Corresponding authors emails: hfan85@scu.edu.cn (HF), wangqy@scu.edu.cn (QW)

## Abstract

Dislocation slip is a general deformation mode and governs the strength of metals. Via discrete dislocation dynamics (DDD) and molecular dynamics (MD) simulations, we investigate the strain rate and dislocation density dependence of the strength of bulk copper single crystals using 192 simulations spanning over 10 orders of magnitude in strain rate $\dot{\varepsilon}$ and 9 orders of magnitude in dislocation density $\rho$. Based on these large set of simulations and theoretical analysis, a new analytical relationship between material strength, dislocation density, strain rate and dislocation mobility is proposed, which is shown to be in excellent agreement with the current simulations as well as with experimental data. The results show that the material strength is a non-monotonic function of dislocation density and displays two universal regimes (first decreasing, then increasing) as the dislocation density increases. The first regime is a result of strain rate hardening, while the second regime is dominated by the classical Taylor forest hardening. Accordingly, the strength displays universally, as a function of strain rate, a rate-independent regime at low strain rates (governed by forest hardening) followed by a rate hardening regime at high strain rates (governed by strain rate hardening). All the results can be captured by a single scaling function, which relates the normalized strength to the coupling parameter ($\rho/\dot{\varepsilon}^{2/3}$) between dislocation density and strain rate. Finally, the fluctuations of dislocation flow are analyzed in terms of the strain rate dependent distribution of dislocation segment velocities. It is found that the fluctuations are governed by another universal scaling function and diverge in the rate independent limit, indicating a critical behavior. The current analysis provides a comprehensive understanding on how collective dislocation motions are governed by the competition between the internal elastic interactions of dislocations, and the stress required to drive dislocation fluxes at a given externally imposed strain rate.

**Keywords:** Discrete dislocation dynamics simulations, strain rate hardening, forest hardening, dislocation plasticity, dislocation kinetics model

## 1. Introduction

Metals are mostly used for their excellent load-bearing capacity, enabled by their mechanical strength and damage tolerance. Serving in practically all engineering fields such as transportation, energy, health, construction and safety, they create an annual global market above 3000 billion Euros[1]. The mechanical properties of most metallic materials exhibit loading rate/time dependency. Particularly many safety-relevant loading scenarios, that metals are subjected to when in service, show significant mechanical response variation with loading rate, for instance, during vehicle crash, metal forming, medical implants or bird strike impact on jet engines. A strain rate hardening response is generic for metallic materials deforming by dislocation slip[2], with exception of a limited regime of deformation conditions in solution-hardened alloys where dislocation-solute interactions may lead to strain rate softening[3,4]. Nevertheless, the relationship between the strain rate and micro-scale deformation mechanisms is still poorly understood, and most dynamic constitutive models (e.g. Johnson-Cook[5], Zerilli-Armstrong[6]) were formulated in a phenomenological or semi-phenomenological manner with several empirical parameters that do not reflect micro-scale deformation mechanisms and need to be fitted to specific experiments with loss of generality[7]. Therefore, it is essential to develop a general understanding of the microscopic mechanisms that control strain rate effects, in order to develop physics-based models that are able to reflect and predict the rate-dependence of the mechanical properties of metals. In BCC (body-centered cubic) metals, such as many steels, rate effects are often related to dislocation core properties (the relatively high atomic-scale Peierls barriers and the associated kink-pair mechanism), which control screw dislocation motion. The resulting temperature and stress dependent mobility of screw dislocations has been incorporated into numerous physics-based plasticity models (see [8,9]). In FCC (face-centered cubic) metals, such as Al and Cu, where dislocation motion is controlled by phonon drag, the situation becomes more complicated because dislocation motion is strongly affected by various collective phenomena related to the mutual elastic interactions among the dislocations. Investigating these phenomena and establishing their rate dependence are the aim of the present study.

Experimental studies on single-crystalline Cu[10], Al[11], and LiF[12] as well as on polycrystalline Cu[13], Al[7] spanning 9 orders of magnitude in strain rate showed that the flow stress exhibits a weakly rate-dependent response at low strain rates followed by a rate hardening response at high strain rates. It has been argued that the rate-independent regime is dominated by dislocation forest interactions and/or dislocation interactions with grain boundaries or precipitates. On the other hand, the rate hardening regime was attributed to viscous drag forces acting on dislocations[7]. In this case, the flow stress acting on dislocations was related to the dislocation velocity through the dislocation drag coefficient, and from dislocation velocity to strain rate through the Orowan relationship. Accordingly, the direct relationship between stress and strain rate depends on the ratio between the drag coefficient and the density of 'mobile' dislocations. This poses serious problems: drag coefficients predicted from rate dependent stress-strain curves under the assumption that all dislocations are mobile are always significantly higher than theoretical estimates, and also higher than drag coefficients deduced from direct velocity measurements[14-16]. Such discrepancy persists even if additional scattering mechanisms beyond viscous phonon drag are considered[17-19]. Kumar et al. conversely used measurements of rate dependent stress-strain curves in conjunction with directly measured drag coefficients to determine mobile dislocation densities, leading to a very low value of the mobile dislocation density at the order of $10^{-5}$ m$^{-2}$ [20]. The problem in all these studies resides in the fact that the mobile dislocation density is not a directly observable quantity. Also, it may be argued that the attribute 'mobile' is somewhat ill-defined since, depending on the loading conditions, any dislocation (including those were temporarily rendered immobile) can become mobile again. This is particularly important when load-path or strain-rate changes are imposed. As a consequence of the conceptual difficulties engendered by introducing the distinct categories of 'mobile' and 'immobile' dislocations, many fundamental questions regarding the relationship between the externally imposed strain rate and the internal collective dynamics of dislocations have never been properly answered. These questions concern not only the relationship between strain rate and average dislocation velocity and its dependence on dislocation density, but also the relationship between individual and collective dislocation behaviors and the meaning of the term 'mobile dislocation density'. To settle these questions, a systematic investigation is required that focuses on the problem: *how dislocations move*.

Discrete dislocation dynamics (DDD) simulations provide *in situ* observations of collective dislocation behavior during plastic flow and can therefore provide fundamental insights into the mechanisms controlling strain rate effects of dislocation mediated plasticity without the need of relying on ad-hoc assumptions. In DDD simulations[21-27], dislocations are coarse-grained as discrete elastic lines and most relevant dislocation mechanisms are accounted for in a physics-based fashion (dislocation glide, cross-slip, multiplication, annihilation, long-range interaction, junction formation, etc.). Over the past two decades, DDD has been extensively employed to investigate various aspects of dislocation mediated plasticity, such as dislocation-dislocation interactions[28-30], dislocation interactions with grain boundaries[31], twin boundaries[32], precipitates[33], and cracks[34]. The two dimensional (2D) DDD approach was previously employed to study dislocation mobility at high strain rates[35], and showed that dislocation inertia effects may be important for the accurate prediction of the dynamical properties of dislocations at high strain rates $> 10^5$ s$^{-1}$ [36]. Using three-dimensional (3D) DDD simulations combined with finite element method, Liu et al. observed that the dislocation patterns change from non-uniform to uniform under high strain rates[37]. Wang et al. performed 3D-DDD simulations and found that while almost all dislocations are mobile at high strain rates[38], a very small percentage of the dislocations move at a speed approaching the shear wave velocity[39]. Under shock loading at super high strain rates, dislocation homogeneous nucleation plays an important role in dynamical plasticity[40,41]. 3D-DDD simulations were also employed to study shock deformation in silicon crystals under laser shock peening, and the dislocation density and dislocation multiplication rate are strongly dependent on the laser processing conditions[42,43]. While DDD simulations were applied to a wide range of problems in dislocation plasticity, the aforementioned fundamental questions pertaining to strain rate dependency have not been systematically investigated. Especially, essential quantities such as the mean dislocation velocity and distribution of dislocation velocity, which are difficult to be determined experimentally, were rarely studied, although they can be naturally obtained from 3D-DDD simulations.

To analyze the strain rate dependence of collective dislocation plasticity, a total of 192 simulations were conducted using primary 3D-DDD and additional molecular dynamics (MD) simulations. In these simulations, the effects of dislocation density (varied over 9 orders of magnitude) and strain rate (varied over 10 orders of magnitude) on the plastic deformation behavior of bulk copper single-crystals were studied. To ensure that the results are not contingent on the simulation method, large scale MD simulations of heavily dislocated samples were conducted additionally and included in the analysis. The mean dislocation velocity and velocity distribution were analyzed in detail and universal characteristics of collective dislocation behavior were revealed. Based on this comprehensive database, we derived a universal analytical relationship between dislocation density, strain rate, material strength and dislocation mobility, which predicts strain rate and dislocation density effects on the plastic properties of metals in terms of a single parameter that combines dislocation density and strain rate.

## 2. Computational method

3D-DDD simulations were performed using the open source code, ParaDiS (v2.5.1), developed at Lawrence Livermore National Laboratory[22]. In ParaDiS, dislocations are discretized into sequences of individual interconnected dislocation segments, each of which carries elastic distortion and associated stress field. Under external applied load $\sigma_{ex}$, each dislocation segment experiences a force per unit length

$$F = b \cdot (\sigma_{ex} + \sigma_{dis}) \times \xi + F_0 + F_{self} \qquad (1)$$

where $\xi$ is the dislocation line direction, $b$ is the Burgers vector of the dislocation segment, $\sigma_{dis}$ is the long-range interaction stress between the current dislocation and others, $F_{self}$ is the dislocation self-force, and $F_0$ is the lattice friction force. Under this total force, each dislocation segment glides on its slip plane. During dislocation glide, short-range dislocation interactions are taken into account, including junction formation and breaking, cross-slip, dislocation annihilation and multiplication. In recent years, ParaDiS was employed frequently to model crystal plasticity in various situations, such as bulk strain hardening, grain boundary strengthening, precipitation hardening and deformation twinning[44-47]. Here, ParaDiS is used to quantify the strain rate effects on collective dislocation behavior in plastically deforming bulk copper (Cu) single-crystals. All DDD simulations were conducted for cubic cells with periodic boundary conditions in three directions. The cube edges are aligned with the three orthogonal crystal lattice directions $X = [100]$, $Y = [010]$, and $Z = [001]$, respectively. To minimize artifacts induced by the periodic boundary conditions, the simulation cell size must be several times larger than the characteristic wavelength of the microstructure (here the dislocation spacing which is estimated as the inverse square root of the dislocation density $\rho$)[48]. Accordingly, the simulation cell size is adjusted according to dislocation density $\rho$, from 1 mm for the lowest dislocation density of $2.3 \times 10^7$ m$^{-2}$ to 100 nm for the highest dislocation density of $1.4 \times 10^{16}$ m$^{-2}$. The material parameters used in all DDD simulations are those of FCC Cu: shear modulus, $G$ = 54.6 GPa; Poisson ratio, $v$ = 0.324; magnitude of Burgers vector, $b$ = 0.25 nm.

In many previous DDD simulations, the initial dislocation configurations consist of Frank-Read dislocation sources (a dislocation ending at two pinning nodes)[32]. Such initial conditions are not only inconsistent with Burgers vector conservation, since the dislocation ends within the crystal, but might also cause artifacts in the dynamics, as the artificially introduced pinning nodes are much stronger than naturally formed ones. Therefore, here we introduced infinite-length dislocations spanning two periodic cells, which are equi-distributed over the 12 possible slip systems. A typical example of the initial configuration is shown in Fig. 1(a). The initial dislocation density, $\rho$, was varied over 9 orders of magnitude ($2.3 \times 10^7$ m$^{-2}$ ~ $1.4 \times 10^{16}$ m$^{-2}$). The initial dislocation configuration was first relaxed under zero stress. During the relaxation, the dislocation density decreases due to dislocation reactions driven by dislocation-related internal stresses. The relaxation is terminated once the incremental plastic strain is less than $10^{-7}$ in 10 $ns$ (~$10^4$ simulation cycles). Figure 1(a) shows that the plastic strain is approaching saturation, indicating that the dislocation configuration approaches a stable state. A representative relaxed dislocation network in the inset of Fig. 1(a) shows a large number of naturally forming dislocation junctions with a very wide spectrum of junction lengths. It should be noted that the accumulated plastic strain produced during the relaxation is significant (up to 0.12% in simulations with a high initial dislocation density). If the relaxation step would be omitted, this accumulative plastic strain would show as a pre-strain occurring during the elastic loading stage. Thus, the relaxation step is important to accurately represent a crystal in equilibrium. Then, a constant strain rate $\dot{\varepsilon}$ is imposed parallel to the simulation cell edge along the Z direction. The imposed strain rate was varied by 7 orders of magnitude from 0.1 s$^{-1}$ to $10^6$ s$^{-1}$. To account for the effect of variations in the initial dislocation network, each simulation was run at least three times (except the case of 0.1 s$^{-1}$) with the same initial dislocation density but different random distributions. A total of 186 simulations were thus performed.

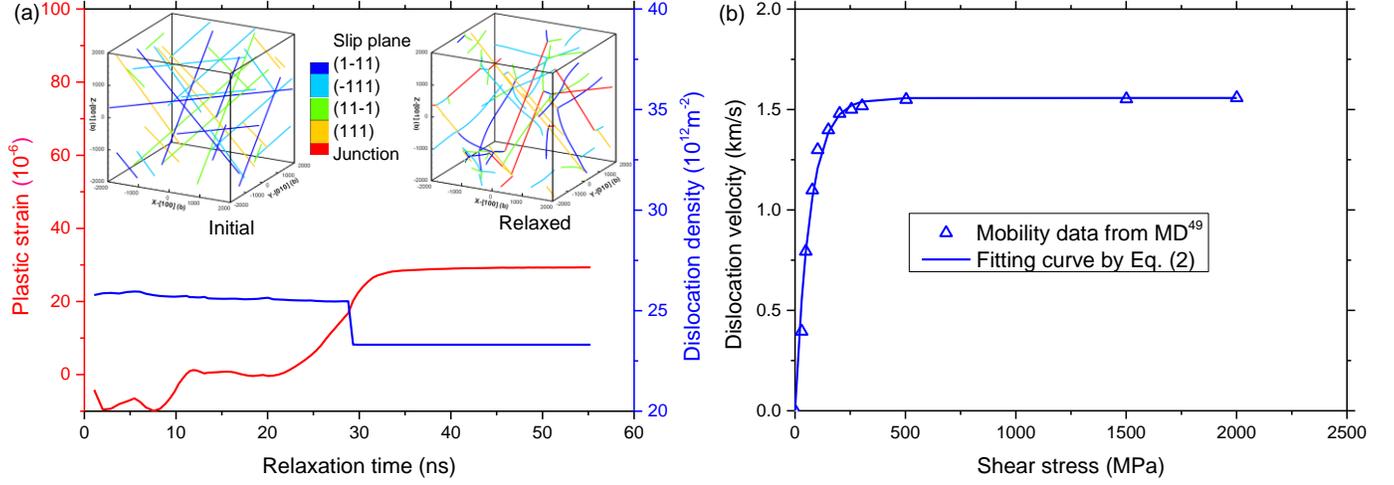

Fig. 1. Plastic strain during relaxation and dislocation mobility law in 3D-DDD simulations. (a) Plastic strain during relaxation for a simulation with an initial dislocation density of $\rho = 2.5 \times 10^{13}$ m$^{-2}$ in bulk copper. Insets show the initial dislocation configuration and relaxed configuration. (b) Dislocation velocity versus resolved shear stress for an edge dislocation as predicted from MD simulations[49], and the exponential dislocation mobility law shown in Eq. (2). Screw dislocation mobility is assumed equal to edge dislocation mobility.

During all the DDD simulations, a fixed strain rate $\dot{\varepsilon}$ is applied uniaxially along Z-direction. The total plastic strain inside the simulation cell is calculated from the area $A$ swept by the dislocation segment, $\varepsilon^{\mathrm{p}} = \sum \frac{A}{2V} (\mathbf{n} \otimes \mathbf{b} + \mathbf{b} \otimes \mathbf{n})$, where $\mathbf{n}$ is the unit normal to the slip plane, $\mathbf{b}$ is the Burgers vector, $V$ is the volume of the simulation cell[50]. Then the response in stress can be obtained by $\sigma = E(\varepsilon - \varepsilon^{\mathrm{p}})$. In high strain rate experiments, the flow stress is believed to be closely related to the mobility of dislocations[14-16]. Thus, to accurately predict dislocation kinetics in high strain rate simulations, a realistic dislocation mobility law is needed. Recent MD simulations of edge dislocation velocity versus resolved shear stress in Cu[49], reproduced in Fig. 1(b), show a non-linear dislocation mobility relationship. Screw dislocation mobility is comparable with edge dislocation mobility. In the current DDD simulations, we utilize an exponential mobility rule of the form

$$v = v_{\mathrm{max}}(1 - \exp(-k\tau)) \qquad (2)$$

where $v_\text{max} = 1.5579$ km/s is the upper limit of the dislocation velocity and $k = 0.0146$ MPa$^{-1}$ is a constant. This mobility law matches reasonably well the MD predictions (see Fig. 1(b)). It is worth noting that the functional form of Eq. (2) also provides a good fit for velocity-stress curves in other FCC metals (e.g. Ni, Al and Al/Mg alloys)[51]. We finally note that this velocity law reduces, in the regime of low to intermediate velocity, to the often used linear drag law, $v = (b/B)\tau$ with linear drag coefficient $B = 1.6 \times 10^{-5}$ Pa·s. At low to intermediate velocity, Eq. (2) represents a linear proportionality between stress and dislocation velocity as expected for drag-controlled dislocation motion. At the same time, the exponential saturation avoids unphysical behavior that would otherwise occur associated with dislocations passing the sound velocity barrier. Comparison with MD simulations, where the inertial and relativistic effects on dislocation motion are naturally included, demonstrates that Eq. (2) provides an adequate representation of collective dislocation behavior even in the high velocity regime (see Fig. 1(b) and our comparison of collective DDD and MD data below).

To ensure that the current predictions are not contingent on simulation method, large scale MD simulations were conducted additionally. The MD simulations were performed using the MD simulation package LAMMPS[52], with the potential for FCC Al[53]. The cubic simulation cell has a size of 113.4 nm with periodic boundary conditions applied in three directions and contains 88 million atoms. In the MD simulation cell, we initially introduced dislocation loops with the same size as the simulation cell[54]. Six initial dislocation densities were considered from $10^{15}$ m$^{-2}$ to $2.8 \times 10^{16}$ m$^{-2}$. After relaxation which was achieved through a conjugate gradient algorithm, a dislocation network forms. Then a strain rate of $2.5 \times 10^8$ s$^{-1}$ was applied on the simulation cell to study the dislocation dynamics during plastic flow.

In all DDD and MD simulations, only dislocation-mediated plasticity has been considered since other deformation modes (e.g. twinning and phase transformation) are only active at shock loading stresses in excess of 10 GPa[55], a regime that is beyond the stresses of interest in this study. Finally, homogenous dislocation nucleation in the crystal was also neglected since previous MD simulations[56] indicate a homogenous nucleation stress of ~10 GPa in copper, which is higher than any yield stresses reached in the current analysis.

## 3. Results and discussions

The resolved shear stresses at yield as obtained from all the DDD and MD simulations are compiled and shown in a double-logarithmic manner in Fig. 2. We define the axial yield stress $\sigma_y$ as the axial stress at a plastic strain of $\varepsilon^p = 0.2\%$, and the resolved shear stress at yield (short: yield stress) as $\tau_y = m\sigma_y$, where $m = 1/\sqrt{6}$ is the Schmidt factor. Figure 2 shows $\tau_y$ as a function of the instantaneous dislocation density $\rho$, and imposed strain rate, $\dot{\varepsilon}$. In Fig. 2(a), it is clearly seen that for a given strain rate the yield stress displays two distinct regimes above and below a critical dislocation density, $\rho_c$. When $\rho < \rho_c$, the yield stress decreases with increasing dislocation density, while for $\rho > \rho_c$, the yield stress increases with increasing density. It is also interesting to note that $\rho_c$ increases with increasing strain rate. In addition, the curves for $\rho > \rho_c$ for all simulated strain rates collapse onto a single line, which coincides with the classical Taylor forest hardening model ($\tau = \alpha G b \sqrt{\rho}$, with $\alpha \approx 0.3$ for FCC metals), showing a strain rate independent response above this critical dislocation density, and the dominance of forest hardening in this regime. On the other hand, while the slopes of the curves for $\rho < \rho_c$ are almost equal, $\tau_y$ increases significantly with increasing strain rate for a given $\rho$, suggesting that in this regime the material is prone to strain rate hardening. In the limit of infinitesimally low strain rate (quasi-static loading), only the second regime remains, indicating that the classical forest hardening mechanism is obtained without the consideration of strain rate effects. Clearly, a competition exists between strain rate hardening and forest hardening. As a result, the material strength is controlled jointly by an internal variable (dislocation density) and an external variable (strain rate). As demonstrated by the black data points in Fig. 2(a), these two regimes are equally observed in DDD and in MD simulations, suggesting that the current predictions are not sensitive to the specific simulation method.

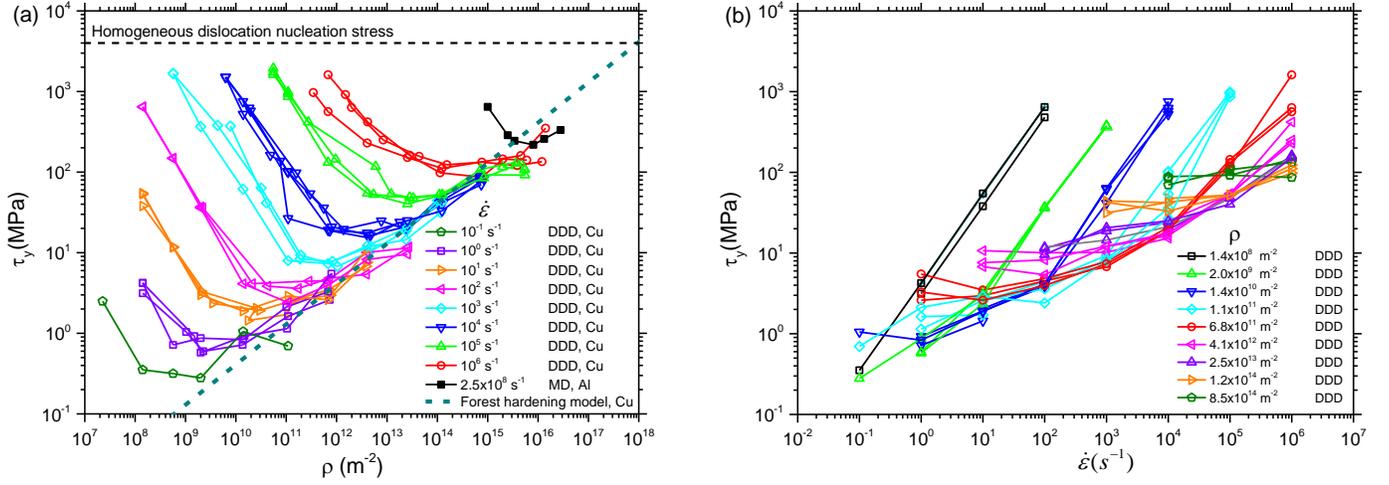

Fig. 2. Yield stress as predicted from current 3D-DDD/MD simulations. Yield stress $\tau_y$ (resolved shear stress at 0.2% plastic strain) as a function of (a) instantaneous dislocation density $\rho$ and (b) strain rate $\dot{\varepsilon}$. In Fig. (b), no MD points are shown because the MD simulations were conducted at only one strain rate.

The yield stress as a function of strain rate as obtained from DDD simulations is shown in Fig. 2(b) for different dislocation densities. No MD data are shown because a wide variation of strain rates is difficult to achieve in MD simulations, which therefore were conducted at a single strain rate only. At first glance, in Fig. 2(b) we see a quite complex picture: For the low dislocation density of $\rho = 1.4 \times 10^8$ m$^{-2}$, $\tau_y$ increases linearly (note the slope is unity in the double-logarithmic plot) with increasing strain rate over the simulated strain rate range. At intermediate dislocation densities, the stress level attained at a given strain rate in the linear regime progressively decreases as the dislocation density increases. At the same time, we observe a cross-over from a linear strain rate dependence at high strain rates towards a low strain rate regime where the slope in the double-logarithmic plot decreases with decreasing strain rate. This cross-over shifts to higher strain rates as dislocation density increases. In the low strain rate regime, the curves approach a horizontal asymptote (rate independent yield stress) with an asymptotic stress level that increases with increasing dislocation density. At the high dislocation density simulated in the DDD approach, viz. $\rho = 8.5 \times 10^{14}$ m$^{-2}$, the yield stress is almost rate independent over the entire range of simulated strain rates. In fact, as we shall demonstrate below, the cross-over from rate independent behavior to a linear rate dependence of the yield stress is a generic feature of the competition between strain rate hardening and forest hardening. Such cross-over also agrees well with extensive experimental observations[10-12,57]. That the cross-over cannot be observed for the low and high dislocation densities probed in our simulations is a consequence of the limited range of strain rates attainable in our DDD simulations.

To analyze the behavior observed in our simulations, we note that the stress rate relates to the strain rate and plastic strain rate through the simple equation

$$\dot{\sigma} = E(\dot{\varepsilon} - \dot{\varepsilon}^{\mathrm{p}}) \quad . \tag{3}$$

We first consider the behavior at extremely high strain rates and/or very low dislocation densities. Since the dislocation velocity cannot exceed the maximum value of $v_{\max}$, for a given dislocation density $\rho$, there exists an absolute upper limit of the plastic strain rate that can be accommodated by dislocation glide. This limit is given by $\dot{\varepsilon}^p_{\max} = m\rho_a b v_{\max}$ (Orowan) where $\rho_a = f_a \rho = 2\rho/3$ is the dislocation density on the active slip systems. If a strain rate $\dot{\varepsilon}$ above this limit is imposed, Eq. (3) has no stationary solution and, hence, the stress is bound to increase indefinitely until, at a stress around 10 GPa, homogeneous dislocation nucleation sets in and the ensuing dramatic dislocation density increase allows to accommodate the imposed strain rate. We denote this scenario as the *exhaustion regime* of the rate dependent response, where the existing dislocations are insufficient to produce the imposed strain rate. Corresponding stress-strain curves are depicted in Fig. 3(a), where the imposed strain rate lies above the strain rate limit for the two lowest dislocation densities (red and green curves in the inset of Fig. 3(a)). As shown in Appendix A, 2% yield stresses in this stage are proportional to the ratio $\dot{\varepsilon}/\rho$.

Next, we move to lower strain rates or higher dislocation densities. Once the imposed strain rate falls below (or dislocation density increases) $\dot{\varepsilon}^p_{\max} = m\rho_a b v_{\max}$, then Eq. (3) possesses a dislocation density dependent quasi-stationary solution, where the stress is implicitly related to the plastic strain rate via $\dot{\varepsilon} = \dot{\varepsilon}^p = m\rho_a b v_m(\tau)$ ($v_\mathrm{m}$ is the mean velocity of dislocations on the active slip systems). In Fig. 3(a), all stress-strain curves with dislocation densities above $\rho = 4.8 \times 10^{10}\,\mathrm{m}^{-2}$ fulfill this condition. These curves are characterized by a sharp transition between an elastic loading stage, and a plastic flow stage where the stress fluctuates around a nearly constant level.

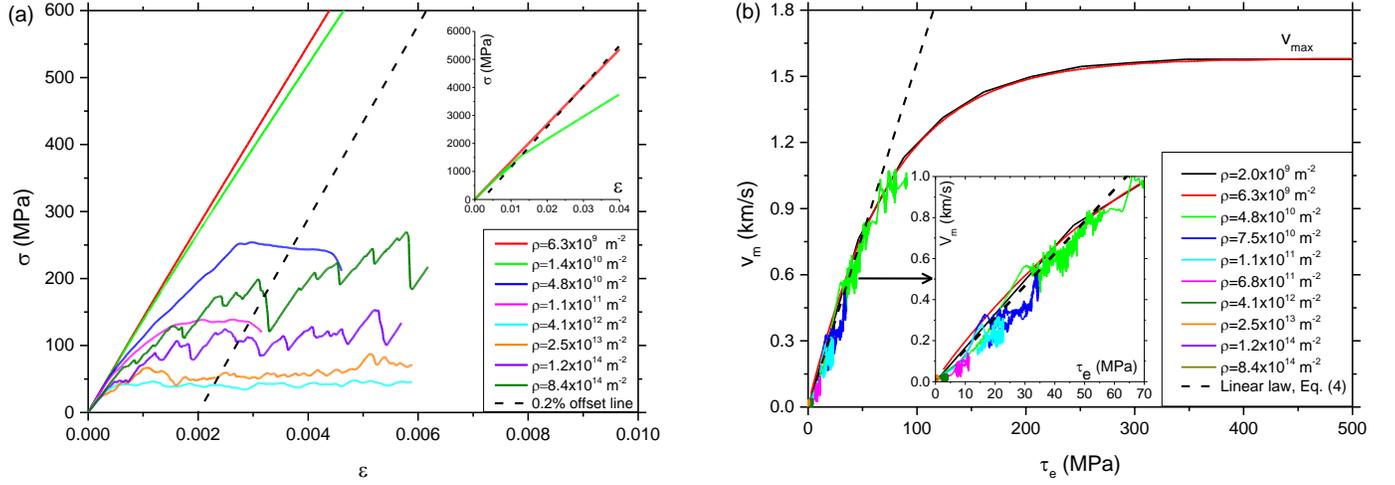

Fig. 3. Stress-strain curves and mean dislocation velocity as predicted from our DDD simulations. (a) Stress-strain curves at applied strain rate of $10^4$ s$^{-1}$ and different dislocation densities. (b) Mean velocity $v_m$ of dislocations on active slip systems versus effective stress $\tau_e$ at applied strain rate of $10^4$ s$^{-1}$ and different dislocation densities.

Within the plastic flow regime, we again first look at the limit of low dislocation densities, where dislocation interactions can be neglected in comparison with the external stress needed to drive the dislocations. This is referred to as the *rate hardening regime* of the rate dependent response. Since all dislocation velocities are well below the maximum velocity, the dislocation mobility law can be linearized, i.e. $v_m = (b/B)\tau$. Hence $\dot{\varepsilon} = f_a m \tau \rho b^2 / B$ and $\tau = B\dot{\varepsilon}/(f_a m \rho b^2)$ which suggests again a linear relationship between the yield stress and the ratio $\dot{\varepsilon}/\rho$. We can see that both the exhaustion regime and the rate hardening regime possess the same dependence of stress on strain rate and dislocation density (see Appendix A for further discussion). The simulation data of Fig. 2 follow this behavior for low dislocation densities or high strain rates.

In the opposite limit of high dislocation density and/or low strain rate, the stress needed to drive dislocations is fully controlled by the mutual interactions of dislocations. In this *Taylor forest hardening regime*, the yield stress of any dislocation arrangement must follow the Taylor relationship, $\tau = \alpha\mu b\sqrt{\rho}$ (see[58] for a general argument regarding this point). This relationship agrees well with the data in Fig. 2 in the regime of high dislocation densities and/or low strain rates.

The next question is whether the three different regimes can be unified into a consistent picture of the strain rate dependence of crystal plasticity. A straightforward idea is that the mean driving stress for dislocation motion is given by an effective stress that equals the resolved shear stress, diminished by the dislocation resistance stress or Taylor stress: $\tau_e = \tau - \alpha\mu b\sqrt{\rho}$. We then expect that the mean dislocation velocity on the active slip systems follows Eq. (2), with the local resolved shear stress replaced by the effective shear stress $\tau_e$. Figure 3(b) shows that the mean dislocation velocity follows well this prediction for a wide range of dislocation densities, as obtained from our 3D-DDD simulations with an applied strain rate of $10^4$ s$^{-1}$. Outwith the exhaustion regime, the dislocation mobility law can be linearized, as shown in the inset of Fig. 3(b). Accordingly,

$$v_m = \tau_e b/B = (\tau - \alpha G b\sqrt{\rho})b/B \tag{4}$$

Using Eqs. (3) and (4) and Orowan's formula, it can be shown that

$$\tau_y = \frac{B\dot{\varepsilon}}{mf_a\rho b^2} + \alpha G b\sqrt{\rho} \tag{5a}$$

which can be alternatively expressed in terms of dimensionless variables to obtain a representation independent of material parameters:

$$T_1 = \frac{\tau_y (mf_a)^{1/3}}{G^{2/3}(B\dot{\varepsilon})^{1/3}} = \frac{1}{P} + \alpha\sqrt{P} \quad , \quad T_2 = \frac{\tau_y}{Gb\sqrt{\rho}} = E + \alpha \tag{5b}$$

where the scaled dislocation density and strain rate are, respectively, given by

$$P = \left(\frac{mf_a Gb^3}{B}\right)^{2/3} \frac{\rho}{\dot{\varepsilon}^{2/3}} \quad , \quad E = P^{-3/2} = \frac{B}{mf_a Gb^3} \frac{\dot{\varepsilon}}{\rho^{3/2}} \tag{5c}$$

Equation (5a) defines a dislocation kinetics model that provides a generic relationship between material strength, dislocation density, strain rate, and the related material parameters like dislocation mobility. This relationship can be stated in the universal forms of Eqs. (5b) and (5c) that are independent of material-specific parameters. As demonstrated in Figs. 4(a, b), these unified models not only allow to collapse all the data in Figs. 2(a) and 2(b) onto two universal curves, but also help to aggregate data obtained for different materials both experimentally and by simulations into the same generic relationship. In particular, experimental data from different materials and for a wide range of deformation conditions follow the same generic curve as the DDD simulation data for Cu, and the same is true for MD simulation data obtained for Al. It is worth noting that the present models even extend to data in the exhaustion regime, as discussed in Appendix A.

In Eqs. (5a-b), the second terms on the right-hand side control the mechanical behavior in the Taylor hardening regime at low strain rates (or high dislocation densities), and the first terms control the behavior in the strain rate hardening regime at high strain rates (or low dislocation densities). The transition between the two regimes can be identified with the minimum of the stress vs. dislocation density curve, which lies at $P = E^{-2/3} = (2/\alpha)^{2/3}$ in a scaled representation. At this minimum, the second term on the right-hand side of Eq. (5b), i.e., the Taylor hardening stress, is exactly twice the rate hardening stress. The absolute values of the critical dislocation density and the minimum stress are given by

$$\rho_c = \left(\frac{2B\dot{\varepsilon}}{\alpha m f_a G b^3}\right)^{2/3} \quad \text{and} \quad \tau_{\min} = \frac{3}{2}\alpha Gb\sqrt{\rho_c} = \left(\frac{27\alpha^2 G^2 B\dot{\varepsilon}}{4 m f_a}\right)^{1/3} \tag{6}$$

$\tau_{\min}$ is the minimum material strength mediated by dislocation plasticity at a given strain rate, which is significant to the community of mechanics and materials. Figure 4(c) shows excellent agreement between the prediction of Eq. (6), the data from current 3D-DDD/MD simulations and published experimental results.

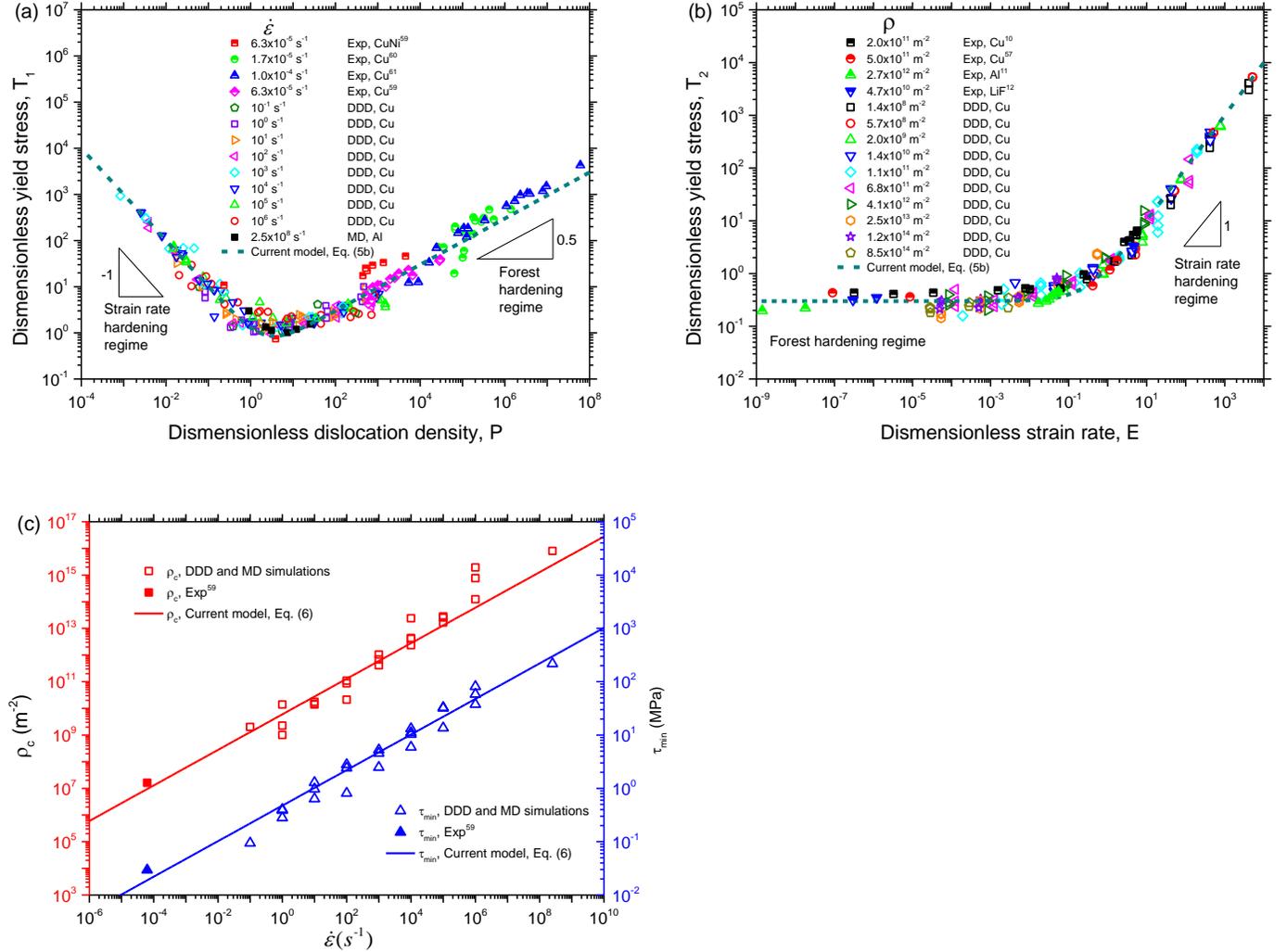

Fig. 4. Comparison between the current models of Eqs. (5, 6), 3D-DDD/MD simulation data and published experimental results[10-12,57,59-61]. (a) Dimensionless yield stress versus dimensionless dislocation density. (b) Dimensionless yield stress versus dimensionless strain rate. (c) The minimum stress $\tau_{\min}$ and critical dislocation density $\rho_c$ at the transition point between forest hardening and rate hardening regimes, as a function of strain rate $\dot{\varepsilon}$.

In many phenomenological plasticity models, the distinction between rate hardening and Taylor hardening terms is absent. Instead, the two-regime response is fitted over a limited range of strain rates by a power law relationship in the form of $\tau_y \propto \dot{\varepsilon}^n$, where $n$ is assumed to represent the strain rate sensitivity of the material. From our analysis it is clear that such a procedure does not adequately represent the intrinsic features of collective dislocation motion. To establish the intrinsic material parameters that control the rate dependency of plastic flow, the forest hardening term should be subtracted from the measured flow stresses such as to produce a linear relationship between strain rate and stress. From this relationship one can determine the coefficient of the strain rate hardening term,

$$s = \frac{B}{m f_a \rho b^2} \qquad (7)$$

which we propose as a physical measure of rate sensitivity in plastic flow of FCC metals. From Eq. (7), the strain rate sensitivity is seen to be mainly controlled by the damping coefficient, $B$, and dislocation density, $\rho$, in a combination, which can explain many corresponding experimental observations in unified form. A higher dislocation density thus is expected to lead to lower strain rate sensitivity. This is in good agreement with experimental observations showing a decrease in strain rate sensitivity with increasing pre-strain[62,63]. Also the strain rate sensitivity increases with increasing temperature for FCC crystals[64] since the dislocation damping coefficient is linearly dependent on temperature[51].

The dimensionless parameters, P and E, in Eq. (5c) not only govern the strain rate and dislocation density dependence of the yield stress, but also control the statistics of dislocation motion. Again, we observe a clear distinction between Taylor hardening and strain rate hardening regimes. This is seen in Fig. 5, which shows the second moment of the dislocation velocity distribution obtained from current DDD simulations, normalized by the square of the mean velocity $\langle v \rangle = f_a v_m$ of all dislocations. In the strain rate hardening regime at low P values (P < 1), the mean square velocity is of the order of the mean velocity squared, i.e.

fluctuations are small in absolute terms and the second moment of the velocity distribution is approximately independent on dislocation density. At high values (P >> 1), the second moment of the velocity distribution grows as $P^{3/2}$. A theoretical expression describing this behaviour can be derived by analysing the microscopic energy dissipation (the work expended in moving dislocations against the drag force) and equating this to the macroscopic dissipated energy (the work expended macroscopically to create a plastic strain). The derivation is given in Appendix B, and the result reads in scaled notation

$$\frac{\langle v^2 \rangle}{\langle v \rangle^2} = \frac{1}{f_a}\left(\alpha P^{3/2} + 1\right) = \frac{1}{f_a}\left(\frac{\alpha}{E} + 1\right) \tag{8}$$

As shown in Fig. 5, this relationship gives a good description of the increase of fluctuations in the regime of high dislocation densities and/or low strain rates that we observe in DDD simulations. Note that the left-hand side can be interpreted as a 'dissipation ratio' where the numerator is proportional to the actual dissipated energy, and the denominator is proportional to the fictitious dissipation in a hypothetical system of non-interacting dislocations of the same density and strain rate.

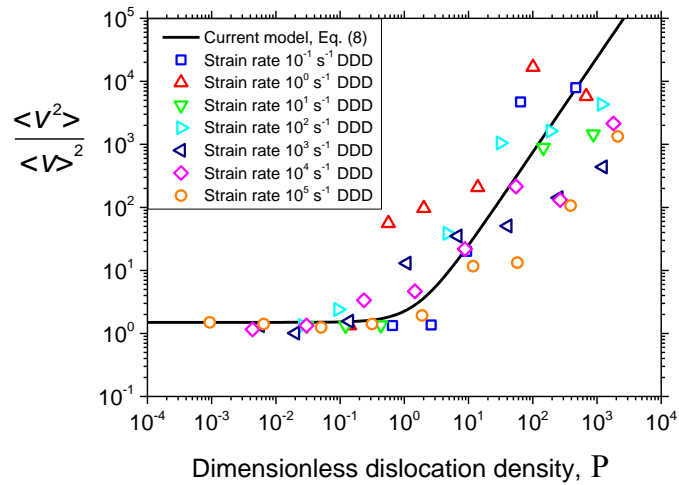

Fig. 5. Squared variation coefficient of the dislocation velocity distribution (dissipation ratio) in 3D-DDD simulations, as a function of the dimensionless dislocation density. Symbols are data sets for different strain rates and full line is the theoretical prediction of Eq. (8).

The dimensionless parameters P and E not only control the magnitude of fluctuations but also govern the statistics of the dislocation velocities: dislocation velocity distributions pertaining to the same P/E values are identical if properly rescaled. This is illustrated in Fig. 6 showing dislocation velocity distributions from current DDD simulations. In the rate hardening regime (Fig. 6(a) for a small value of P), we find bimodal distributions with one peak at near zero velocity which represents dislocations on inactive slip systems, and one peak at high velocity comprising all dislocations on active slip systems. In the latter case, the velocities of these dislocations are fairly uniform and scatter around the peak velocity value that is required to produce the imposed strain rate. From Eqs. (4) and (5a)

$$v_{\text{peak2}} \approx v_m = \frac{\langle v \rangle}{f_a}, \text{ so } \frac{v_{\text{peak2}}}{\langle v \rangle} \approx \frac{1}{f_a} \tag{9}$$

The so unified picture that emerges in the rate hardening regime is thus clear: we can distinguish immobile dislocations, which are the dislocations on the inactive slip systems from mobile dislocations, which comprise all dislocations on the active slip systems. These dislocations move at the velocity needed to produce the imposed strain rate, with only minor velocity fluctuations. The flow stress is dictated by the drag on dislocations, and dislocation-dislocation interactions are fairly unimportant.

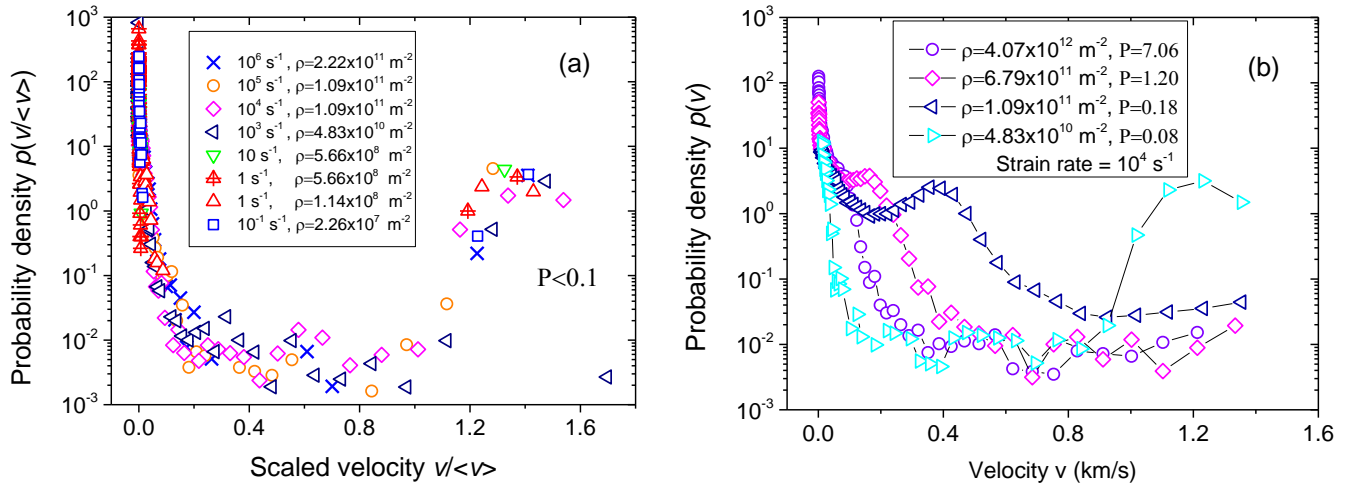

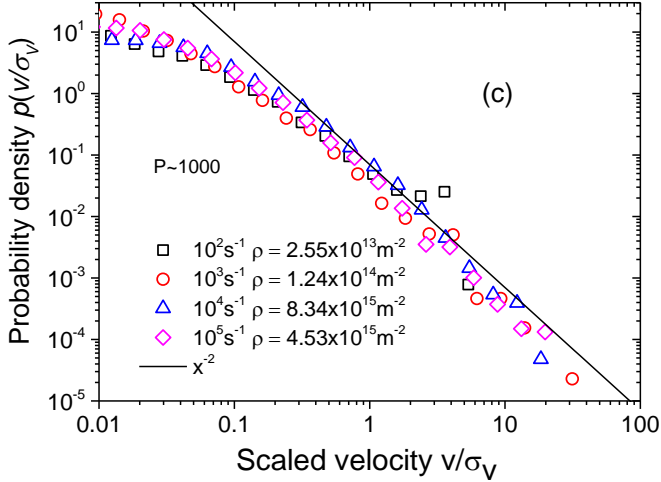

Fig. 6. Probability distributions of dislocation velocities as observed in 3D-DDD simulations. (a) Distributions in the strain rate hardening regime, P < 0.1; (b) distributions in the intermediate regime 0.1 < P < 10; (c) distributions in the Taylor forest hardening regime, P > 1000, in double logarithmic representation, where the full line indicates a slope of -2. $\sigma_V$ is the standard deviation of dislocation velocity distribution, and <v> is the mean velocity of all dislocations.

As the dimensionless strain rate parameter E decreases or the density parameter P increases towards unity, the high-velocity peak of the velocity distribution shifts to lower values and ultimately merges, for high P, with the low-velocity peak (see Fig. 6(b)), leading to unimodal dislocation velocity distribution that is typical and characteristic of the Taylor hardening regime. In the rate independent limit P → ∞ the velocity distribution acquires scale free features as the probability density $p(v)$ decreases, for high velocity, in inverse proportion with $v^2$, leading to the diverging fluctuations.

## 4. Conclusions

In this work, the strain rate dependence of collective dislocation dynamics was studied using a large set of 3D-DDD (discrete dislocation dynamics) and MD (molecular dynamics) simulations, spanning nine orders of magnitude in dislocation density and ten orders of magnitude in strain rate. The performed 192 simulations indicate that the material strength displays two regimes, a strain rate hardening regime where the yield stress increases in proportion with strain rate and in inverse proportion with dislocation density, and a regime of classical Taylor hardening where the yield stress is approximately rate independent and follows the Taylor relationship. All results can be described in terms of a scaled dislocation density P or strain rate E, which combine dislocation density, strain rate, and material parameters in such a manner that the corresponding yield stress can be expressed through a universal material independent relationship. The analytical relationship describes not only our simulations over the entire range of strain rates and dislocation densities, but also a wide range of experimental data published in the literature. The scaled dislocation density/strain rate parameters not only control the yield stress but also govern the statistics of dislocation velocities. In the rate-hardening regime of high strain rates/low dislocation densities, we find bimodal velocity distributions, where dislocations on inactive slip systems remain immobile whereas dislocations on active slip systems move with small fluctuations in a quasi-laminar manner at the velocity needed to match the imposed strain rate. In the Taylor hardening regime, on the other hand, the velocity distributions have scale free characteristics and decrease monotonically towards high velocity according to a $p(v) \propto v^{-2}$ power law.

The current results have far-reaching consequences both regarding the interpretation of experiments and the constitutive modeling of crystal plasticity. The interpretation of experiments that try to probe the strain rate dependence of dislocation motion and to establish drag coefficients has hinged in the idea that it is possible to distinguish a mobile dislocation density, which moves at a velocity that is dictated by the externally applied stress, and an immobile dislocation density that consists of dislocations remaining essentially stationary. Our investigation demonstrates that such a distinction makes sense only in the rate hardening regime of very high strain rates and/or low dislocation densities. Only experiments conducted in this regime can yield results that are amenable to direct interpretation. However, most actual experiments have been conducted at low strain rates and/or high dislocation densities, i.e. in the Taylor hardening regime (see Fig. 4(a)). In these cases, identifying the mobile dislocation density with the dislocation density on the active slip systems is bound to systematically over-estimate drag effects, and the introduction of a mobile fraction of the dislocation density is tantamount to introducing a variable that cannot be determined independently either by experiments or, as our study demonstrates, in simulations. At the same time, our results offer a way out of this dilemma, as we provide a universal yield stress relationship, which contains only the total dislocation density and strain rate (both measurable quantities) together with material parameters. One of these parameters is the poorly documented drag coefficient (as discussed in the introduction), and the remaining parameters (shear modulus, Burgers vector) are accurately known. Thus, by re-scaling experimental data obtained from samples at different strain rates to fall on the master curve provided by our Eq. (5) and depicted in Fig. 4, it is possible to determine drag coefficient $B$ without the need to rely on assumptions regarding a spurious mobile dislocation density.

Regarding constitutive modeling, we note that, starting from the Kubin-Estrin model[65], dislocation based crystal plasticity models regularly contain a 'mobile dislocation density' as a constitutive variable (for recent examples, see[66]). Our analysis demonstrates that, in the Taylor hardening regime, the distribution of dislocation segment velocities offers no means to define such a quantity in a meaningful manner. In the strain rate hardening regime, by contrast, its definition is straightforward but trivial since the mobile dislocation density simply encompasses all dislocations on active slip systems. This is very problematic from a conceptual point of view, in particular since to our knowledge there has been no experimental determination of the same quantity, which would require large-scale *in situ* experiments to be conducted with single-dislocation resolution and sufficient statistical sampling of the microstructure.

In summary, our investigation provides a unifying picture of the strain rate and dislocation density dependence of collective dislocation dynamics over a so far unprecedented range of scales. In the regime of comparatively low strain rates or high dislocation densities, in which most laboratory experiments are conducted, collective dynamics of dislocations appears as a highly turbulent flow process. Once a sufficiently high applied stress causes the dislocation arrangement to lose metastability, complex relaxation processes lead to highly irregular dynamics with a scale free dislocation velocity spectrum.

## Acknowledgements

This work was supported by National Natural Science Foundation of China (U1730106, 11672193), Alexander von Humboldt Foundation, Chinese State Administration of Foreign Experts Affairs #MS2016XNJT044 and US National Science Foundation Award #DMR-1609533.

## Author contributions

HF designed the DDD and MD simulations; HF, JAE, QW performed computer simulations; HF and MZ deduced the models; HF, JAE, QW, MZ, DR provided partial financial supports and research requisites; All authors contributed to discussions of the results and revisions of the manuscript and approved the final version.

## Competing financial interests

All authors declare no competing interests.

## Appendix A:

As mentioned above, Eq. (5a) was derived on the basis of two simplifications: low dislocation velocity and negligible strain hardening rate, so that the mobility law can be linearized. As shown in Fig. 3(a), these simplifications are reasonable for the cases of intermediate and high dislocation densities. Here, we discuss the cases of low dislocation densities in exhaustion regime, which exhibits high dislocation velocities approaching the maximum velocity (i.e. $v_m \approx v_{\max}$) and high strain hardening rates. From Eq. (3) and Orowan's formula, the strain hardening rate is

$$\theta = E(1 - mf_a\rho b v_{\max}/\dot{\varepsilon}) \tag{A1}$$

The yield stress at 0.2% plastic strain is

$$\tau_y = \frac{0.002m}{1/\theta - 1/E} \tag{A2}$$

Eliminating $\theta$ from Eqs. (A1) and (A2) results in

$$\tau_y = 0.002mE\left(\frac{\dot{\varepsilon}}{mf_a\rho b v_{\max}} - 1\right) \tag{A3}$$

Thus, the yield stress is linear with $\dot{\varepsilon}/\rho$, which is similar to the first term in Eq. (5a) (the second term in Eq. (5a) is negligible at low dislocation density). The ratio of the two slopes in Eqs. (A3) and (5a) is $0.002mbE/Bv_{\max}$. As shown in the theory of dislocations[67], the damping coefficient can be expressed as $B = \eta Gb/v_{\max}$, where $\eta \approx 0.002$. The ratio of the two slopes becomes $2m(1+\nu)$, which is comparable to unity, suggesting that Eq. (5a) can still be used at low dislocation densities. That is why Eq. (5a) shows good agreement with the DDD/MD simulations and experiments even in exhaustion regime.

**Appendix B: A fluctuation-dissipation theorem for dislocation plasticity**

**B1. Macroscopic dissipation**

It is a standard assumption in continuum plasticity theory, which is motivated by thermodynamic arguments, that the work expended in creating plastic deformation is entirely dissipated into heat. Hence, the dissipated work per unit volume is equal to the plastic work

$$\frac{dW^{\text{diss}}}{dt} = \frac{dW^{\text{p}}}{dt} = \sigma \dot{\varepsilon}^p \tag{B1}$$

For dislocation plasticity, this statement needs to be qualified: strain hardening is associated with the change of an internal variable (the dislocation density) and, since dislocations carry elastic energy in form of stress and strain fields and the much smaller contribution from the dislocation core energy, this leads to a stored internal energy contribution (a stored defect energy[68]). We estimate the defect energy storage rate as

$$\frac{dE^{\text{def}}}{dt} = E_L \frac{d\rho}{d\varepsilon} \dot{\varepsilon}^p \tag{B2}$$

where $E_L$ is the dislocation line energy. To obtain an upper estimate of the dislocation storage rate, we use the well-established Kocks-Mecking model and note that an upper bound to the dislocation storage rate is obtained by neglecting dynamic recovery. In that case, following[69] and using $E_L = \eta G b^2$, we can estimate

$$\left(\frac{\alpha G b}{2}\right)^2 \frac{\partial \rho}{\partial \varepsilon^p} = \sigma \frac{\partial \sigma}{\partial \varepsilon^p} \leq \sigma \theta_{II}, \quad \frac{dE^{def}}{dt} \leq \frac{\theta_{II}}{G}\left(\frac{4\eta}{\alpha^2}\right)\sigma\dot{\varepsilon}^p = \frac{\theta_{II}}{G}\left(\frac{4\eta}{\alpha^2}\right)\frac{dW^p}{dt} \tag{B3}$$

where $\theta_{II}$ is the initial hardening slope in the limit of athermal rate-independent deformation (in FCC single crystals: hardening stage II). The numerical factor in the bracket on the right-hand side of Eq. (B3) is of the order of 1. Since the hardening slope is only a tiny fraction of the shear modulus ($\theta_{II} \approx G/200$, see ref. [69]), the defect energy storage rate is only a tiny fraction of the plastic work rate. Hence Eq. (B1) holds for dislocation plasticity with only minor corrections due to stored defect energy.

**B2. Microscopic dissipation**

On the dislocation level, dissipation occurs because the work that is expended in moving dislocations is dissipated into the phonon system. The situation is particularly simple for non-relativistic over-damped dislocation motion, because there, due to absence of inertia, all work is instantaneously dissipated. Using $v = (b/B)\tau$ and $F^{PK} = b\tau$ we write the dissipated energy as

$$\frac{dW^{diss}}{dt} = \frac{1}{V}\int_{(\mathbb{C}_V)} F^{PK} v\, ds = \frac{B}{V}\int_{(\mathbb{C}_V)} v^2\, ds \tag{B4}$$

where the integral runs over the set $\mathbb{C}_V$ of all dislocation lines in the volume $V$ considered. Introducing the dislocation density and the mean square dislocation velocity as

$$\rho = \frac{1}{V} \int_{(\mathbb{C}_V)} ds \quad , \quad \langle v^2 \rangle = \frac{\int_{(\mathbb{C}_V)} v^2 ds}{\int_{(\mathbb{C}_V)} ds} \tag{B5}$$

we thus find that the mean velocity square is related to the microscopically dissipated work by

$$\frac{dW^{diss}}{dt} = B\rho \langle v^2 \rangle \tag{B6}$$

**B3. Fluctuation-dissipation relationship**

It is clear that the microscopically dissipated work and the macroscopic dissipated work must be identical. We therefore obtain a relationship between the 'macroscopic' quantities in (B1) and the 'microscopic' quantities in (B6):

$$\frac{dW^{diss}}{dt} = B\rho \langle v^2 \rangle = \sigma \dot{\varepsilon}^p \tag{B7}$$

We write the plastic strain rate now in terms of microscopic quantities (segment velocities) as

$$\dot{\varepsilon}^p = \frac{b}{V} \int_{(\mathbb{C}_V)} m(s)v(s)ds =: m\rho b \langle v \rangle \quad , \quad \langle v \rangle = \frac{\int_{(\mathbb{C}_V)} m(s)v ds}{m \int_{(\mathbb{C}_V)} ds} \tag{B8}$$

where we note that the motion of dislocations on inactive slip systems (Schmidt factor: $m(s)=0$) does not contribute to the $m$-weighted average velocity. Accordingly, this average velocity $\langle v \rangle$ of all dislocations relates to the mean velocity $v_m$ on the active slip systems via $\langle v \rangle = f_a v_m$.

We now use Eq. (5a) of the main paper to write $\sigma = m^{-1}(\alpha Gb\sqrt{\rho} + (1/f_a)(B/b)\langle v \rangle)$. This leads to our final result,

$$\frac{\langle v^2 \rangle}{\langle v \rangle^2} = \frac{\alpha Gb^3}{B}\left(\frac{\rho}{\dot{\varepsilon}^{2/3}}\right)^{3/2} + \frac{1}{f_a} = \frac{1}{f_a}\left(\alpha P^{3/2} + 1\right) \tag{B9}$$

We can envisage Eq. (B9) as an expression for the magnitude of fluctuations. We define the weighted coefficient of variation of dislocation velocities as

$$\text{COV}_M = \frac{\langle \delta v^2 \rangle^{0.5}}{\langle v \rangle} = \left( \frac{\langle v^2 \rangle}{\langle v \rangle^2} - 1 \right)^{1/2} = \frac{\alpha G b^3}{B} \left( \frac{\rho}{\dot{\varepsilon}^{2/3}} \right)^{3/4} \quad \text{(B10)}$$

This quantity measures the magnitude of dislocation velocity fluctuations. As we move to zero or very small strain rates, this quantity diverges which indicates critical behavior. We note that the general idea of the above derivation goes back to Hähner[70] and the case of a linear drag law has, in embryonic form, been previously considered by one of the present authors[71].

## References

1  Mining & Metals in a Sustainable World 2050. *World Economic Forum Report* (2015).
2  Kolsky, H. An Investigation of the Mechanical Properties of Materials at very High Rates of Loading. *P. Roy. Soc. Lond. B* **62**, 676 (1949).
3  Zaiser, M. & Hähner, P. A unified description of strain-rate softening instabilities. *Materials Science and Engineering: A* **238**, 399-406 (1997).
4  Zaiser, M., Glazov, M., Lalli, L. A. & Richmond, O. On the relations between strain and strain-rate softening phenomena in some metallic materials: a computational study. *Comp. Mater. Sci.* **15**, 0-49 (1999).
5  Johnson, G. R. & Cook, W. H. Fracture characteristics of three metals subjected to various strains, strain rates, temperatures and pressures. *Eng. Fract. Mech.* **21**, 31-48 (1985).
6  Zerilli, F. J. & Armstrong, R. W. Dislocation‐mechanics‐based constitutive relations for material dynamics calculations. *J. Appl. Phys.* **61**, 1816-1825 (1987).
7  Salvado, F. C., Teixeira-Dias, F., Walley, S. M., Lea, L. J. & Cardoso, J. B. A review on the strain rate dependency of the dynamic viscoplastic response of FCC metals. *Prog. Mater. Sci.* **88**, 186-231 (2017).
8  Ma, A., Roters, F. & Raabe, D. A dislocation density based constitutive law for BCC materials in crystal plasticity FEM. *Comp. Mater. Sci.* **39**, 91-95 (2007).
9  Khan, A. S. & Liang, R. Behaviors of three BCC metal over a wide range of strain rates and temperatures: experiments and modeling. *Int. J. Plast.* **15**, 1089-1109 (1999).
10  Edington, J. W. The influence of strain rate on the mechanical properties and dislocation substructure in deformed copper single crystals. *The Philosophical Magazine: A Journal of Theoretical Experimental and Applied Physics* **19**, 1189-1206 (1969).
11  Chiem, C. Y. & Duffy, J. Strain rate history effects and observations of dislocation substructure in aluminum single crystals following dynamic deformation. *Mater. Sci. Eng.* **57**, 233-247 (1983).

12	Chiem, C. Y. & Duffy, J. Strain rate history effects in LiF single crystals during dynamic loading in shear. *Mater. Sci. Eng.* **48**, 207-222 (1981).

13	Lea, L. J. & Jardine, A. P. Characterisation of high rate plasticity in the uniaxial deformation of high purity copper at elevated temperatures. *Int. J. Plast.* **102**, 41-52 (2018).

14	Clifton, R. J. Dynamic Plasticity. *J. Appl. Mech.* **50**, 941-952 (1983).

15	Ferguson, W. G., Kumar, A. & Dorn, J. E. Dislocation Damping in Aluminum at High Strain Rates. *J. Appl. Phys.* **38**, 1863-1869 (1967).

16	Armstrong, R. W., Arnold, W. & Zerilli, F. J. Dislocation mechanics of copper and iron in high rate deformation tests. *J. Appl. Phys.* **105**, 023511 (2009).

17	Ferguson, W. G., Hauser, F. E. & Dorn, J. E. Dislocation damping in zinc single crystals. *Br. J. Appl. Phys.* **18**, 411 (1967).

18	Kumar, A., Hauser, F. E. & Dorn, J. E. Viscous drag on dislocations in aluminum at high strain rates. *Acta Metall.* **16**, 1189-1197 (1968).

19	Victoria, M. P., Dharan, C. K. H., Hauser, F. E. & Dorn, J. E. Dislocation Damping at High Strain Rates in Aluminum and Aluminum‐Copper Alloy. *J. Appl. Phys.* **41**, 674-677 (1970).

20	Kumar, A. & Kumble, R. G. Viscous Drag on Dislocations at High Strain Rates in Copper. *J. Appl. Phys.* **40**, 3475-3480 (1969).

21	Van der Giessen, E. & Needleman, A. Discrete dislocation plasticity: a simple planar model *Model. Simul. Mater. Sci. Eng.* **3**, 689-735 (1995).

22	Arsenlis, A. *et al.* Enabling strain hardening simulations with dislocation dynamics. *Model. Simul. Mater. Sci. Eng.* **15**, 553-595. (2007).

23	Kubin, L. P. *et al.* Dislocation microstructures and plastic flow: A 3D simulation. *Solid State Phenom.* **23 & 24**, 455-472 (1992).

24	Ghoniem, N. M., Tong, S. H. & Sun, L. Z. Parametric dislocation dynamics: A thermodynamics-based approach to investigations of mesoscopic plastic deformation. *Phys. Rev. B* **61**, 913-927 (2000).

25	Benzerga, A. A., Bréchet, Y., Needleman, A. & Van der Giessen, E. Incorporating three-dimensional mechanisms into two-dimensional dislocation dynamics. *Model. Simul. Mater. Sci. Eng.* **12**, 159 (2004).

26	Liu, Z. L., Liu, X. M., Zhuang, Z. & You, X. C. A multi-scale computational model of crystal plasticity at submicron-to-nanometer scales. *Int. J. Plast.* **25**, 1436-1455 (2009).

27	Zbib, H. M. & Diaz de la Rubia, T. A multiscale model of plasticity. *Int. J. Plast.* **18**, 1133-1163 (2002).

28	El-Awady, J. A. Unravelling the physics of size-dependent dislocation-mediated plasticity. *Nat. Commun.* **6** (2015).

29	Gao, Y. *et al.* Investigations of pipe-diffusion-based dislocation climb by discrete dislocation dynamics. *Int. J. Plast.* **27**, 1055-1071 (2011).

30	Liu, F. X., Liu, Z. L., Pei, X. Y., Hu, J. Q. & Zhuang, Z. Modeling high temperature anneal hardening in Au submicron pillar by developing coupled dislocation glide-climb model. *Int. J. Plast.* **99**, 102-119 (2017).

31  Fan, H., Aubry, S., Arsenlis, A. & El-Awady, J. A. Orientation influence on grain size effects in ultrafine-grained magnesium. *Scripta Mater.* **97**, 25-28 (2015).

32  Fan, H., Aubry, S., Arsenlis, A. & El-Awady, J. A. The role of twinning deformation on the hardening response of polycrystalline magnesium from discrete dislocation dynamics simulations. *Acta Mater.* **92**, 126-139 (2015).

33  Záležák, T., Svoboda, J. & Dlouhý, A. High temperature dislocation processes in precipitation hardened crystals investigated by a 3D discrete dislocation dynamics. *Int. J. Plast.* **97**, 1-23 (2017).

34  Huang, M., Tong, J. & Li, Z. A study of fatigue crack tip characteristics using discrete dislocation dynamics. *Int. J. Plast.* **54**, 229-246 (2014).

35  Roos, A., De Hosson, J. T. M. & Van der Giessen, E. High-speed dislocations in high strain-rate deformations. *Computational Materials Science* **20**, 19-27 (2001).

36  Wang, Z. Q., Beyerlein, I. J. & Lesar, R. Dislocation motion in high strain-rate deformation. *Philosophical Magazine* **87**, 2263-2279 (2007).

37  Liu, Z. L., You, X. C. & Zhuang, Z. A mesoscale investigation of strain rate effect on dynamic deformation of single-crystal copper. *Int. J. Solids Struct.* **45**, 3674-3687 (2008).

38  Wang, Z. Q., Beyerlein, I. J. & LeSar, R. Plastic anisotropy in fcc single crystals in high rate deformation. *Int. J. Plast.* **25**, 26-48 (2009).

39  Shehadeh, M. A., Zbib, H. M. & Diaz De La Rubia, T. Modelling the dynamic deformation and patterning in fcc single crystals at high strain rates: dislocation dynamics plasticity analysis. *Philos. Mag.* **85**, 1667-1685 (2005).

40  Shehadeh, M. A., Bringa, E. M., Zbib, H. M., McNaney, J. M. & Remington, B. A. Simulation of shock-induced plasticity including homogeneous and heterogeneous dislocation nucleations. *Applied Physics Letters* **89**, 171918 (2006).

41  Gurrutxaga-Lerma, B., Balint, D. S., Dini, D., Eakins, D. E. & Sutton, A. P. Attenuation of the Dynamic Yield Point of Shocked Aluminum Using Elastodynamic Simulations of Dislocation Dynamics. *Phys. Rev. Lett.* **114**, 174301 (2015).

42  Cheng, G. J. & Shehadeh, M. A. Multiscale dislocation dynamics analyses of laser shock peening in silicon single crystals. *Int. J. Plast.* **22**, 2171-2194 (2006).

43  Cheng, G. J. & Shehadeh, M. A. Dislocation behavior in silicon crystal induced by laser shock peening: A multiscale simulation approach. *Scripta Materialia* **53**, 1013-1018 (2005).

44  Hussein, A. M., Rao, S. I., Uchic, M. D., Dimiduk, D. M. & El-Awady, J. A. Microstructurally based cross-slip mechanisms and their effects on dislocation microstructure evolution in fcc crystals. *Acta Mater.* **85**, 180-190 (2015).

45  Fan, H., Li, Z., Huang, M. & Zhang, X. Thickness effects in polycrystalline thin films: Surface constraint versus interior constraint. *International Journal of Solids and Structures* **48**, 1754-1766 (2011).
29

46   Fan, H., Aubry, S., Arsenlis, A. & El-Awady, J. A. Grain size effects on dislocation and twinning mediated plasticity in magnesium. *Scripta Mater.* **112**, 50-53 (2016).

47   Lehtinen, A., Granberg, F., Laurson, L., Nordlund, K. & Alava, M. J. Multiscale modeling of dislocation-precipitate interactions in Fe: From molecular dynamics to discrete dislocations. *Phys. Rev. E* **93**, 013309 (2016).

48   de Souza, O. N. & Ornstein, R. L. Effect of periodic box size on aqueous molecular dynamics simulation of a DNA dodecamer with particle-mesh Ewald method. *Biophys. J.* **72**, 2395-2397 (1997).

49   Oren, E., Yahel, E. & Makov, G. Dislocation kinematics: a molecular dynamics study in Cu. *Model. Simul. Mater. Sci. Eng.* **25**, 025002 (2017).

50   Akarapu, S., Zbib, H. M. & Bahr, D. F. Analysis of heterogeneous deformation and dislocation dynamics in single crystal micropillars under compression. *Int. J. Plasticity* **26**, 239-257 (2010).

51   Olmsted, D. L. & et al. Atomistic simulations of dislocation mobility in Al, Ni and Al/Mg alloys. *Model. Simul. Mater. Sci. Eng.* **13**, 371 (2005).

52   Plimpton, S. Fast Parallel Algorithms for Short-Range Molecular Dynamics. *J. Comp. Phys.* **117**, 1-19 (1995).

53   Zope, R. R. & Mishin, Y. Interatomic potentials for atomistic simulations of the Ti-Al system. *Physical Review B* **68**, 024102 (2003).

54   Zepeda-Ruiz, L. A., Stukowski, A., Oppelstrup, T. & Bulatov, V. V. Probing the limits of metal plasticity with molecular dynamics simulations. *Nature* **550**, 492 (2017).

55   Meyers, M. A. *Dynamic Behavior of Materials*. (Wiley, 1994).

56   Tschopp, M. A. & et al. Atomistic simulations of homogeneous dislocation nucleation in single crystal copper. *Model. Simul. Mater. Sci. Eng.* **15**, 693 (2007).

57   Stelly, M. & Dormeval, R. in *High Velocity Deformation of Solids.* (eds Kozo Kawata & Jumpei Shioiri) 82-97 (Springer Berlin Heidelberg).

58   Zaiser, M. & Sandfeld, S. Scaling properties of dislocation simulations in the similitude regime. *Modelling and Simulation in Materials Science and Engineering* **22**, 065012 (2014).

59   Hildebrand, H. The effect of the initial dislocation density on dislocation multiplication and work-hardening characteristics of copper single crystals. *Phys. Status Solidi A* **12**, 239-249 (1972).

60   Livingston, J. D. The density and distribution of dislocations in deformed copper crystals. *Acta Metall.* **10**, 229-239 (1962).

61   Van Drunen, G. & Saimoto, S. Deformation and recovery of [001] oriented copper crystals. *Acta Metall.* **19**, 213-221 (1971).

62   Campbell, J. D. & Ferguson, W. G. The temperature and strain-rate dependence of the shear strength of mild steel. *Philos. Mag.* **21**, 63-82 (1970).

63   Patrick, L., Annette, B., Kirsten, D. & Wolfgang, B. Influence of Strain Rate, Temperature, Plastic Strain, and Microstructure on the Strain Rate Sensitivity of Automotive Sheet Steels. *Steel Res. Int.* **84**, 426-442 (2013).

64  Altan, T. & Boulger, F. W. Flow Stress of Metals and Its Application in Metal Forming Analyses. *Journal of Engineering for Industry* **95**, 1009-1019 (1973).

65  Estrin, Y. & Kubin, L. P. Local strain hardening and nonuniformity of plastic deformation. *Acta Metall.* **34**, 2455-2464 (1986).

66  Roters, F. *et al.* Overview of constitutive laws, kinematics, homogenization and multiscale methods in crystal plasticity finite-element modeling: Theory, experiments, applications. *Acta Mater.* **58**, 1152-1211 (2010).

67  Hirth, J. P. & Lothe, J. *Theory of Dislocations*. 2nd edn, (John Wiley and Sons, 1982).

68  Zaiser, M. Local density approximation for the energy functional of three-dimensional dislocation systems. *Phys. Rev. B* **92**, 174120 (2015).

69  Mecking, H. & Kocks, U. F. Kinetics of flow and strain-hardening. *Acta Metall.* **29**, 1865-1875 (1981).

70  Hähner, P. On the foundations of stochastic dislocation dynamics. *Appl. Phys. A* **62**, 473-481 (1996).

71  Zaiser, M. Statistical modelling of dislocation systems. *Mater. Sci. Eng. A* **309-310**, 304-315 (2001).